\documentclass[useAMS,usenatbib,usegraphicx]{mn2e}

\newcommand{\der}{{\rm d}} 
 \newcommand{\mnras}{MNRAS}
\newcommand{\apj}{ApJ} \newcommand{\apjl}{ApJ} 
 \newcommand{\apjs}{ApJS} 
\newcommand{\aap}{A\&A}

 \newcommand{\h}{_{\rm h}} \newcommand{\m}{_{\rm m}}
 
\newcommand{\R}{{R_{\rm f}}} 
 
\newcommand{\cc}{_{\rm ci}}  
\newcommand{\co}{_{\rm c}} \newcommand{\cz}{_{\rm c0}}
\newcommand{\pko}{_{\rm m}} \newcommand{\pkz}{_{\rm m0}} 
\newcommand{\tru}{_{\rm p}}

\newcommand{\pk}{_{\rm mi}} \newcommand{\nest}{^{\rm nn}}
 \newcommand{\ES}{_{\rm es}}

  \newcommand{\eff}{}

\newcommand{\F}{^{\rm th}} 
 
\newcommand{\p}{_{\rm p}}  
 
\newcommand{\modot}{M$_\odot$\ } \newcommand{\modotc}{M$_\odot$}
\newcommand{\ti}{t_{\rm i}} 
\newcommand{\ii}{_{\rm i}}

\newcommand{\beq}{\begin{equation}} \newcommand{\eeq}{\end{equation}}
 \newcommand{\beqa}{\begin{eqnarray}}
\newcommand{\eeqa}{\end{eqnarray}} \newcommand{\lav}{\langle}
\newcommand{\rav}{\rangle}  
\newcommand{\vir}{_{\rm vir}}

\begin{document}

\title[Halo Mass Function] {Halo Mass Definition and Multiplicity Function}

\author[Juan et al.]{Enric Juan\thanks{E-mail:
    ejrovira@am.ub.es}, Eduard Salvador-Sol\'e, Guillem
  Dom\`enech and Alberto Manrique \\Institut de Ci\`encies del Cosmos.
  Universitat de Barcelona, UB--IEEC.  Mart{\'\i} i Franqu\`es 1,
  E-08028 Barcelona, Spain}


\maketitle
\begin{abstract}
Comparing the excursion set and CUSP formalisms for the derivation of
the halo mass function, we investigate the role of the mass definition
in the properties of the multiplicity function of cold dark matter
(CDM) haloes. We show that the density profile for haloes formed from
triaxial peaks that undergo ellipsoidal collapse and virialisation is
such that the ratio between the mean inner density and the outer local
density is essentially independent of mass. This causes that, for
suited values of the spherical overdensity $\Delta$ and the linking
length $b$, SO and FoF masses are essentially equivalent to each other
and the respective multiplicity functions are essentially the
same. The overdensity for haloes having undergone ellipsoidal collapse
is the same as if they had formed according to the spherical top-hat
model, which leads to a value of $b$ corresponding to the usual virial
overdensity, $\Delta\vir$, equal to $\sim 0.2$. The multiplicity
function resulting from such mass definitions, expressed as a function
of the top-hat height for spherical collapse, is very approximately
universal in all CDM cosmologies. The reason for this is that, for
such mass definitions, the top-hat density contrast for ellipsoidal
collapse and virialisation is close to a universal value, equal to
$\sim 0.9$ times the usual top-hat density contrast for spherical
collapse.
\end{abstract}



\section{INTRODUCTION}\label{intro}

Large scale structure harbours important cosmological
information. However, such a fundamental property as the halo mass
function (MF) is not well-established yet. Besides the lack of an
accurate description of non-linear evolution of density fluctuations,
there is the uncertainty arising from the fact that the boundary of a
virialised halo is a fuzzy concept. As a consequence, the halo mass
function depends on the particular mass definition adopted, its shape
being only known in a few cases and over limited mass and redshift
ranges.

The various halo mass definitions found in the literature arise from
the different halo finders used in simulations
\citep{Kea11}. For instance, in the Spherical Overdensity
(SO) definition \citep{LC94}, the mass of a halo at the time $t$ is
that leading to a total mean density $\bar \rho\h(R\h)$ equal to a
fixed, constant or time-varying, overdensity $\Delta$ times the mean
cosmic density $\bar\rho(t)$,
\beq
\bar\rho\h(R\h)=\Delta \bar\rho(t)\,.
\label{first}
\eeq
While in the Friends-of-Friends (FoF) definition \citep{Dea85}, the
mass of a halo is the total mass of its particles, identified by means
of a percolation algorithm with fixed linking length $b$ in units of
the mean inter-particle separation.

The main drawback of the FoF definition is that, for large values of
$b$, it tends to over-link haloes. Its main advantage is that it can
be applied without caring about the symmetry and dynamical state of
haloes. Haloes are, indeed, triaxial rather than spherically
symmetric, harbour substantial substructure and may be undergoing a
merger, which complicates the use of the SO definition. However, one
can focus on virialised objects and consider the spherically averaged
density profile $\rho\h(r)$ and mass profile $M(r)$ around the
peak-density, in which case the FoF mass coincides with the mass
inside the radius $R\h$ where spheres of radius $b$ harbour two
particles in average \citep{LC94},
\beq
\rho\h(R)=\frac{3}{2\pi}\,b^{-3}\bar\rho(t)\,.
\label{second}
\eeq

Equations (\ref{first}) and (\ref{second}) imply the relation
\beq
\Delta=\frac{3 F(c)}{2\pi}\,b^{-3}\,,
\label{cor}
\eeq
between $\Delta$ and $b$ for haloes of a given mass $M$, where $F(c)\equiv
\bar\rho\h(R\h)/\rho\h(R\h)$ is a function of halo concentration $c$.

As $c$ depends on $M$, there is no pair of $\Delta $ and
$b$ values satisfying equation (\ref{cor}) for all $M$ at the same
time. Consequently, there is strictly no equivalent SO and FoF mass
definitions \citep{Mea11}. Yet, numerical simulations show that, at
least in the Standard Cold Dark Matter (SCDM) cosmology, FoF masses
with $b=0.2$, from now on simply FoF(0.2), tightly correlate with SO
masses with overdensity $\Delta$ equal to the so-called virial value,
$\Delta\vir\approx 178$, from now on SO($\Delta\vir$)
\citep{CL96}. This correlation is often interpreted as due to the fact
that haloes are close to isothermal spheres, for which $F(c)$ is
equal to 3, so equation (\ref{cor}) for $b=0.2$ implies
$\Delta\approx 178$.

Simulations also show that, in any cold dark matter (CDM) cosmology,
FoF(0.2) haloes have a multiplicity function that, expressed as a
function of the top-hat height for spherical collapse, is
approximately universal \citep{Jea00,Wh02,Wea06,Lea07,Tea08,Cea10} and
very similar to that found for SO($\Delta\vir$) haloes
\citep{Jea00,MW02}. As $\Delta\vir$ may substantially deviate from 178
depending on the cosmology, such a similarity cannot be due to the
roughly isothermal structure of haloes as suggested by the SCDM
case. Moreover, the universality of this multiplicity function is hard
to reconcile with the dependence on cosmology of halo density profile
\citep{Coea11}. On the other hand, haloes do not form through
spherical collapse but through ellipsoidal collapse. For all theses
reasons, the origin of such properties is unknown. Having a reliable
theoretical model of the halo MF would be very useful for trying to
clarify these issues.

Assuming the spherical collapse of halo seeds, \citet{PS} derived a MF
that is in fair agreement with the results of numerical simulations
(e.g. \citealt{Eea88,WEF93,LC94,bm}), although with substantial
deviations at both mass ends
\citep{LC94,Gea98,Jea01,Wh02,Ree03,Hea06}. An outstanding
characteristic of the associated multiplicity function is its
universal shape as a function of the height of density
fluctuations. Whether this characteristic is connected with the
approximately universal multiplicity function of simulated haloes for
FoF(0.2) masses is however hard to tell.

Bond et al.~(1991) re-derived this MF making use of the so-called
excursion set formalism in order to correct for cloud-in-cloud
(nested) configurations. This formalism was adopted in subsequent
refinements carried out with the aim to account for the more realistic
ellipsoidal collapse \citep{M95,LS98,ST02}. The excursion set
formalism has also recently been modified \citep{Pea12,ESP} to account
for the fact that density maxima (peaks) in the initial density field
are the most probable halo seeds (\citealt{HP13}).

In an alternative approach, the extension to peaks was directly
attempted from the original Press-Schechter MF
\citep{B89,CLM89,PH90,AJ90,bm,H01}. The most rigorous derivation along
this line was by Manrique and Salvador-Sol\'e (1995,
hereafter MSS; see also \citealt{MSS98}), who applied the so-called
{\it ConflUent System of Peak trajectories} (CUSP) formalism, based on
the Ansatz suggested by spherical collapse that ``there is a
one-to-one correspondence between haloes and non-nested peaks''.

A common feature of all these derivations is that they
assume monolithic collapse or pure accretion. While in hierarchical
cosmologies there are certainly periods in which haloes evolve by
accretion, major mergers are also frequent and cannot be neglected. We
will comeback to this point at the end of the paper. A second and more
important issue in connection with the problem mentioned above is that
none of these theoretical MFs makes any explicit statement on the halo
mass definition presumed, so the specific empirical MF they are to be
compared with is unknown.

Recently, Juan et al.~(2013, hereafter JSDM) have shown that,
combining the CUSP formalism with the exact follow-up of ellipsoidal
collapse and virialisation developed by Salvador-Sol\'e et al.~(2012,
hereafter SVMS), it is possible to derive a MF that adapts to any desired halo
mass definition and is in excellent agreement with the results of
simulations.

In the present paper, we use the excursion set and CUSP formalisms to
explain the origin of the observed properties of the halo multiplicity
function.

In Section \ref{approaches}, we recall the two different approaches
for the derivation of the MF. In Section \ref{defs}, we investigate
the mass definition implicitly assumed in such approaches. The
origin of the similarity of the multiplicity function for FoF(0.2) and
SO($\Delta\vir$) masses and of its approximate universality is
addressed in Sections \ref{simil} and \ref{univ}, respectively. Our
results are discussed and summarised in Section \ref{discuss}.

All the quantitative results given throughout the paper are for the
concordant $\Lambda$CDM cosmology with $\Omega_{\Lambda}=0.73$,
$\Omega_{\rm m}=0.23$, $\Omega_{\rm b}=0.045$, $H_0=0.71$ km s$^{-1}$
Mpc$^{-1}$, $\sigma_8=0.81$, $n_{\rm s}=1$ and Bardeen et al. (1986,
hereafter BBKS) CDM spectrum with \citet{S95} shape parameter.

\section{Mass Function}\label{approaches}

All derivations of the halo MF proceed by first identifying the seeds
of haloes with mass $M$ at the time $t$ in the density field at an arbitrary
small enough cosmic time $\ti$ and then counting those seeds.

\subsection{The Excursion Set Formalism}

In this approach, halo seeds are assumed to be spherical overdense
regions in the initial density field smoothed with a {\it top-hat}
filter that undergo spherical collapse. 

The time of spherical collapse (neglecting shell-crossing) of a seed
depends only on its density contrast, so there is a one-to-one
correspondence between haloes with $M$ at $t$ and density
perturbations with fixed density contrast $\delta\cc$ at the filtering
radii $\R$ satisfying the relations
\beq
\delta\cc(t)=\delta\co(t) \frac{D(\ti)}{D(t)}\,
\label{delta1}
\eeq
\beq
\R(M)=\left[\frac{3M}{4\pi\bar\rho\ii}\right]^{1/3}\,.
\label{Rf1}
\eeq
In equations (\ref{delta1}) and (\ref{Rf1}), $\bar\rho\ii$ is the mean
cosmic density at $t=\ti$, $\delta\co(t)$ is the almost universal
density contrast for spherical collapse at $t$ linearly extrapolated
to that time and $D(t)$ is the cosmic growth factor. In the
Einstein-de Sitter universe, $D(t)$ is equal to the cosmic scale
factor $a(t)$ and $\delta\co(t)$ is equal to $3(12\pi)^{2/3}/20\approx
1.686$. While, in the concordant model and the present time $t_0$, 
$D(t_0)$ is a factor 0.760 smaller than $a(t_0)$ and $\delta\co(t_0)$
is equal to $\approx 1.674$ (e.g. \citealt{H00}).

Equation (\ref{Rf1}) is valid to leading order in the perturbation,
the exact relation between $\R$ and $M$ being
\beq
\R(M,t)=\left\{\frac{3M}{4\pi\bar\rho\ii[1+\delta\cc(t)]}\right\}^{1/3}\,.
\label{Rf2}
\eeq
The interest of adopting the approx relation (\ref{Rf1}) is that the
filtering radius then depends only on $M$. This greatly simplifies the
mathematical treatment.

Following \citet{PS}, every region with density contrast greater than
or equal to $\delta\cc(t)$ at the scale $\R(M)$ will give rise at $t$
to a halo with mass greater than or equal to $M$. Consequently, the MF,
i.e. the comoving number density of haloes per infinitesimal mass
around $M$ at $t$, is simply the $M$-derivative of the volume fraction
occupied by those regions, equal in Gaussian random density fields to
\beq 
V(M,t)=\frac{1}{2}\,{\rm erfc}\left[\frac{1}{\sqrt{2}}\,\frac{\delta\cc(t)}{\sigma_0\F(M,\ti)}\right]\,,
\label{vol}
\eeq
divided by the volume
$M/\bar\rho(t)$ of one single seed, 
\beq 
\frac{\partial n_{\rm PS}(M,t)}{\partial M}=\frac{\bar\rho(t)}{M}\,\frac{\partial
  V(M,t)}{\partial M}\,.
\label{PSvol}
\eeq
In equation (\ref{vol}), $\sigma_0\F(M,\ti)$ is the top-hat rms
density fluctuation of scale $M$ at $\ti$. 

But this derivation does not take into account that overdense regions
of a given scale may lie within larger scale overdense regions, which
translates into a wrong normalisation\footnote{The normalisation
  condition reflects the fact that all the matter in the universe must
  be in the form of virialised haloes.} of the MF
(\ref{vol})--(\ref{PSvol}). To correct for this effect, \citet{BCEK}
introduced the excursion set formalism. The density contrast $\delta$
at any fixed point tends to decrease as the smoothing radius $\R$
increases, so, using a sharp k-space filter, $\delta$ traces a
Brownian random walk, easy to monitor statistically. In particular,
one can estimate the number of haloes reaching $M$ at $t$ by counting
the excursion sets $\delta(R)$ intersecting $\delta\cc(t)$ at any
scale $\R(M)$. The important novelty of this approach is that,
whenever a halo undergoes a major merger, $\delta$ increases instead
of decreasing, so every trajectory $\delta(R)$ can intersect
$\delta\cc(t)$ at more than one radius $R$, meaning that there will be
haloes appearing within other more massive ones. Therefore, to correct
for cloud-in-cloud configurations, one must simply count the excursion
sets intersecting $\delta\cc(t)$ for the first time as $R$ decreases
from infinity (or $\sigma_0\F$ increases from zero), as if they were
absorbed at such a barrier. The MF so obtained has identical form as
the Press-Schechter one (eqs.~[\ref{PSvol}]-[\ref{vol}]) but with an
{\it additional factor two}, 
\beq \frac{\partial n\ES(M,t)}{\partial M}=2\frac{\partial n_{\rm
    PS}(M,t)}{\partial M}\,,
\label{esMF}
\eeq
yielding the right normalisation of the excursion set MF.

Note that, as the height of a density fluctuation, defined as the
density contrast normalised to the rms value at the same scale, is
constant with time, the volume $V(M,t)$ (eq.~[\ref{vol}]) can be
written as a function of
$\nu\ES=\delta\cc(t)/\sigma_0\F(M,\ti)=\delta\co(t)/\sigma_0\F(M,t)=\delta\cz(t)/\sigma_0\F$,
where $\delta\cz(t)$ is $\delta\co(t)D(t_0)/D(t)$ and $\sigma_0\F$
stands for the 0th order spectral moment at the current time
$t_0$. Thus, the resulting MF (eq.~[\ref{PSvol}]) is independent of
the arbitrary initial time $\ti$.

But the assumptions made in this derivation are not fully
satisfactory: i) Every overdense region does not collapse into a
distinct halo; only those around peaks do. Unfortunately, the
extension of the excursion set formalism to peaks is not trivial
\citep{Pea12,ESP}; ii) Real haloes (and peaks) are not spherically
symmetric but triaxial, so halo seeds do not undergo spherical but
ellipsoidal collapse. Unfortunately, the implementation in this
approach of ellipsoidal collapse is hard to achieve due to the
dependence on $M$ of the corresponding critical density contrast
\citep{ST02,Pea12,ESP}. iii) The formation of haloes involves not only
the collapse of the seed, but also the virialisation (through
shell-crossing) of the system, which is hard to account for. And iv)
there is a slight inconsistency between the top-hat filter used to
monitor the dynamics of collapse and the sharp k-space window used to
correct for nesting. The use of the top-hat filter with this latter
purpose is again hard to implement due to the correlation between
fluctuations at different scales in top-hat smoothing \citep{MP12}.

\subsection{The CUSP Formalism}\label{CUSP}

In this approach, halo seeds are regions around {\it non-nested
  (triaxial) peaks} that undergo {\it ellipsoidal collapse and
  virialisation}. The use of a Gaussian filter is mandatory in this
case for the reasons given below.

The time of collapse and virialisation of triaxial seeds depends not
only on their density contrast $\delta$ at the suited scale $R$, like
in spherical collapse, but also on their ellipticity and density slope
(e.g. \citealt{Peebles}). However, all peaks with given $\delta$ at
$R$ have similar ellipticities and density slopes
(JSDM). Consequently, all haloes at $t$ can be seen to arise from
non-nested triaxial peaks at $\ti$ with the same fixed density
contrast $\delta\pk$ at any filtering radius $\R$. Then, taking
advantage of the freedom in the boundary of virialised haloes, we can
adopt at $t$ a suited mass definition so to exactly match the
one-parameter family of resulting halo masses $M(\R,t)$. By doing
this, we will end up with a one-to-one correspondence between haloes
with $M$ at $t$ and non-nested (triaxial) peaks at $\ti$ with
$\delta\pk$ at $\R$, according to the relations
\beq 
\delta\pk(t)=\delta\pko(t) \frac{D(\ti)}{D(t)}\,
\label{deltat}
\eeq
\beq
\R(M,t)=\frac{1}{q(M,t)}\left[\frac{3M}{4\pi\bar\rho\ii}\right]^{1/3}\,.
\label{rm}
\eeq

Equations (\ref{deltat})--(\ref{rm}) are very similar to equations
(\ref{delta1})--(\ref{Rf1}). However, the pair of functions
$\delta\pko(t)$ and $q(M,t)$ are now arbitrary, fixing one particular
halo mass definition each (and conversely; see
Sec.~\ref{JSDMcase}). For this reason the subindex ``c'' for
``collapse'' in the density contrasts appearing in the relation
(\ref{delta1}) has been replaced by the subindex ``m'' indicating that
the masses of haloes at $t$ ``match'' a certain definition. The only
constraints these two functions must fulfil, for consistency with the
mass growth of haloes, are: $\delta\pk(t)$ must be a decreasing
function of time and $\R(M,t)$ must be an increasing function of
mass. That consistency condition is precisely what makes the use of
the Gaussian filter mandatory. Such a filter is indeed the only one
guaranteeing, through the relation $R\nabla^2 \delta=\partial
\delta/\partial R$, that the density contrast $\delta$ of peaks
necessarily decreases as the filtering radius $R$ increases, in
agreement with the evolution of halo masses.

The radius $R\p$ of the spherically averaged seed (or protohalo) with
$M$ is now different from the Gaussian filtering radius $\R$. It is
instead equal to $q\tru\R$, with $q\p$ satisfying the relation
\beq
\R(M,t)= \frac{1}{q\tru(M,t)}
\left\{\frac{3M}{4\pi\bar\rho\ii[1+\delta\pk\F(M,t)]}\right\}^{1/3},
\label{rmtru}
\eeq
where $\delta\pk\F(M,t)$ is the density contrast of the peak when
the density field is smoothed with a {\it top-hat} filter encompassing
the same mass $M$, related to the (unconvolved) spherically averaged
density contrast profile of the seed $\delta\p(r)$ through
\beq
\delta\pk\F(M,t)= \frac{3}{R\p^3}\int_0^{R\p} \der r\,r^2\,\delta\p(r)\,.
\label{delta}
\eeq
But factors $[1+\delta\pk\F(M,t)]^{1/3}$ and $q\tru(M,t)$ on the right
of equation (\ref{rmtru}) can be absorbed in the function
$q(M,t)$. Thus, contrarily to the relation (\ref{Rf1}), equation
(\ref{rm}) is exact. Note that $q(M,t)$ is then, to leading order in
the perturbation as in equation (\ref{Rf1}), the radius $R\p$ of the
seed in units of the Gaussian filtering radius $\R$.

We emphasise that, owing to the Gaussian smoothing, the radius of the
filter $\R$ depends, in the CUSP formalism, not only on $M$ but also
on $t$ through $q(M,t)$, while $\delta\pk$ and, hence, $\delta\pko$
are still functions of $t$ alone. This latter functionality may seem
contradictory with the fact that, as pointed out by \citet{ST02} and
recently checked with simulations (e.g.,
\citealt{Rea09,Eea12,Dea13,HP13}), the density contrast for
ellipsoidal collapse depends on the mass of the perturbation. There is
however no contradiction. The CUSP formalism uses a Gaussian filter
instead of a top-hat filter like in all these works and this
difference is crucial because of the freedom introduced by
$q(M,t)$. In top-hat smoothing, $q$ is fixed to one, so considering
ellipsoidal collapse necessarily translates into a value of
$\delta\co$ dependent on $M$ in addition to $t$. While, in Gaussian
smoothing, we can impose that the density contrast for collapse is
independent of $M$ and let $q$ depend on $M$ and $t$.  Note that, when
the density field at $\ti$ is smoothed with a top-hat filter, the
density contrast of halo seeds, $\delta\pk\F(M,t)$, is indeed a
function of $M$ and $t$ in general. The exact way $\delta\pk\F(M,t)$
depends on $M$ will depend, of course, on the mass definition used.

We are now ready to calculate the MF in this approach. Given the
one-to-one correspondence between haloes and non-nested peaks, the
counting of haloes with $M$ at $t$ reduces to count non-nested
peaks with that scale at $\ti$. In the original version of the CUSP
formalism (MSS), such a counting did not take into account the
correlation between peaks at different scales. However, the more
accurate version later developed \citep{MSS98} yielded essentially the
same result, so we will follow here that simple version (see the
Appendix for the more accurate one). For simplicity, we will omit
hereafter any subindex in the Gaussian rms density fluctuation
$\sigma_0$ and in the CUSP height $\nu\equiv
\delta\pk/\sigma_0(M,\ti)=\delta\pko(t)/\sigma_0(M,t)=\delta\pkz(t)/
\sigma_0$, where $\delta\pkz(t)$ is $\delta\pko(t)D(t_0)/D(t)$ and
$\sigma_0$ stands for the 0th order spectral moment at $t_0$. The
subindexes ``th'' and ``es'' in the excursion set counterparts are
enough to tell between the two sets of variables.

The number density of peaks with $\delta\pk$ per infinitesimal $\ln
\sigma_0^{-1}(M,\ti)$ at $\ti$ or, equivalently, with $\delta\pkz$ per
infinitesimal $\ln \sigma_0^{-1}$ at $t_0$ can be readily calculated
from the density of peaks per infinitesimal height around $\nu$,
derived by BBKS. The result is
\beq
N(\sigma_0,\delta\pkz)
=\frac{\lav x\rav\!(\sigma_0,\delta\pkz)}{(2\pi)^2\,R_\star^3\,\gamma}\,\,
{\rm e}^{-\frac{\nu^2}{2}}\,,
\label{npeak}
\eeq
where $R_\star$ and $\gamma$ are respectively defined as
$\sqrt{3}\sigma_1/\sigma_2$ and $\sigma_1^2/(\sigma_0\sigma_2)$, being
$\sigma_j$ the $j$-th order (Gaussian) spectral moment, and $\left\lav
x\right\rav\!(\sigma_0,\delta\pkz)$ is the average curvature
(i.e. minus the Laplacian scaled to the mean value $\sigma_2$) of
peaks with $\delta\pkz$ and $\sigma_0$, well-fitted by the analytic
expression (BBKS)
\beq 
\lav x\rav(\nu)=\gamma
\nu+\frac{3(1-\gamma^2)+(1.216-0.9\gamma^4){\rm
    e}^{-\frac{\gamma}{2}\,\left(\frac{\gamma\nu}{2}\right)^2}}{\left[3(1-\gamma^2)+0.45+(\gamma\nu/2)^2\right]^{1/2}+\gamma\nu/2}\,.
\label{averx}
\eeq

But this number density is not enough for our purposes because we are
interested in counting {\it non-nested} peaks only. The homologous number
density of non-nested peaks, $N\nest(\sigma_0,\delta\pkz)$, can be
obtained by solving the Volterra integral equation
\beqa
N\nest(\sigma_0,\delta\pkz)=N(\sigma_0,\delta\pkz)-\!\!\int_{\ln \sigma_0^{-1}}^\infty \der \ln {\sigma'_0}^{-1}
\nonumber~~~~~~~\\
\times N(\sigma_0,\delta\pkz|\sigma'_0,\delta\pkz)\frac{M(\sigma'_0,\delta\pkz)}{\bar\rho}N\nest(\sigma'_0,\delta\pkz),
\label{nnp}
\eeqa
where the second term on the right gives the density of peaks with
$\delta\pkz$ per infinitesimal $\ln \sigma_0^{-1}$ nested into peaks
with identical density contrast at larger scales, $\ln
{\sigma'_0}^{-1}$. The conditional number density
$N(\sigma_0,\delta\pkz|\sigma'_0,\delta\pkz)$ of peaks with
$\delta\pkz$ per infinitesimal $\ln \sigma_0^{-1}$ subject to lying in
backgrounds with $\delta\pkz$ at $\sigma'_0<\sigma_0$ can also be
calculated from the conditional number density per infinitesimal $\nu$
in backgrounds with $\nu'$ derived by BBKS. The result is
\beq
N(\sigma_0,\delta\pkz|\sigma'_0,\delta\pkz)=\frac{\lav
  x\rav(\tilde\sigma_0,\delta\pkz)}{(2\pi)^2\,R_\star^3\,\gamma\,\sqrt{1 - \epsilon^2}}\, {\rm
  e}^{-\frac{(\nu -\epsilon\,\nu')^2}{2(1-\epsilon^2)}}\,,
\label{int2}
\eeq
where $\nu'$ and $\epsilon$ are respectively defined as
$\delta\pkz/\sigma_0'$ and $\sigma_0^2(R\m)/[\sigma_0\sigma'_0]$,
being $R\m^2$ equal to the arithmetic mean of the squared filtering
radii corresponding to $\sigma_0$ and $\sigma_0'$, and where $\lav
x\rav\!(\tilde \sigma_0, \delta\pkz)$ takes the same form
(\ref{averx}) as $\lav x\rav\!(\sigma_0, \delta\pkz)$ in equation
(\ref{npeak}) but as a function of $\tilde \gamma\tilde \nu$ instead
of $\gamma\nu$, being
\beq 
\tilde\gamma^2=
\gamma^2\left[1 + \epsilon^2\,{(1 - r_1)^2 \over 1 -
    \epsilon^2}\right]
\label{tgamma2}
\eeq
\beq
\tilde\nu(r)\!=\!{\gamma \over \tilde\gamma}\,{1 - r_1 \over 1 -
\epsilon^2}\left[\nu\!\left({1 - \epsilon^2r_1 \over 1 - r_1}\!
\right)\!-\epsilon\nu'\right]\!,
\label{tildbis2}
\eeq
with $r_1$ equal to $[\sigma_0(\R)\sigma_1(R\m)/(\sigma_1(\R)\sigma_0(R\m))]^2$.

Thus, the MF of haloes at $t$ is then
\beq
\frac{\partial n_{\rm CUSP}(M,t)}{\partial M}=\,N\nest[\sigma_0,\delta\pkz]\,\frac{\partial \ln\sigma_0^{-1}}{\partial M}\,.
\label{mf}
\eeq
Note that this expression of the MF is also independent of the
(arbitrary) initial time $\ti$.

The CUSP formalism thus solves all the problems met in the excursion
set formalism: it deals with triaxial peaks that undergo ellipsoidal
collapse and virialisation, conveniently corrected for nesting, and
the smoothing of the initial density field is always carried out with
the same Gaussian filter. The only drawback of this approach is the
need to solve the Volterra equation (\ref{nnp}), which prevents from
having an analytic expression for the resulting MF.

\section{Implicit Halo Mass Definition}\label{defs}

For any theoretical MF to be complete, the mass definition it refers
to must be specified. In other words, one must state the condition
defining the total radius $R\h$ or, equivalently, the spherically
averaged density profile for haloes with different masses at $t$
that result from the specific halo seeds and dynamics of collapse assumed.

\subsection{The Excursion Set Formalism}\label{PScase}

In the excursion set formalism, halo seeds are arbitrary overdense
regions with no definite inner structure, so their typical (mean)
density and peculiar velocity fields are uniform. As a consequence,
the density distribution in the corresponding final virialised objects
is also uniform\footnote{As shown in SVMS, what causes the outwards
  decreasing density profile of virialised objects is the fact that,
  for seeds with outwards decreasing density profiles, virialisation
  progresses from the centre of the system outwards. In the case of
  homogeneous spheres in Hubble expansion, all the shells cross at the
  same time at the origin of the system, so the final object does not
  have an outwards decreasing density profile.}. In addition, the
system is supposed to undergo spherical collapse. Therefore, halo
formation is according to the simple spherical top-hat model, in which
case the typical radii $R\h$ of haloes with different masses at $t$
can be readily inferred \citep{Peebles}.

The virial relation $2T+W=0$ holding for the final uniform
object\footnote{The effects of the cosmological constant at halo
  scales can be neglected.} together with energy
conservation\footnote{In the top-hat spherical model, energy cannot be
  evacuated outwards like in the virialisation of haloes formed
  by the collapse of seeds with outwards decreasing density
  profiles (SVMS), so the total energy is conserved.} imply that $R\h$ is
half the radius of the uniform system at turnaround. This leads to
\beq 
R\h=-\frac{3GM^2}{10 E\p(M)}\,,
\label{mr2}
\eeq
where $E\p(M)$ is the (conserved) total energy of the protohalo with
mass $M$. Taking into account that $E\p$ is, to leading order in the
perturbation, equal to $-\delta\cc(t) GM^2/\R$ (see
eqs.~[\ref{E2}]--[\ref{dM}] for $\rho\p=\bar\rho\ii[1+\delta\cc(t)]$),
equation (\ref{mr2}) takes the form
\beq
R\h=\left[\frac{3M}{4\pi\Delta\vir(t) \bar\rho(t)}\right]^{1/3}\,
\label{r}
\eeq
or, equivalently,
\beq
\bar\rho\h(R\h)=\Delta\vir(t)\, \bar\rho(t)\,,
\label{Delta}
\eeq
where we have introduced the so-called virial overdensity
corresponding to the spherical top-hat model,
\beq
\Delta\vir(t)\equiv 
\left[\frac{10\,\delta\co(t)a(t)}{3\,D(t)}\right]^3\,.
\label{Deltavir}
\eeq

Comparing equations (\ref{first}) and (\ref{Delta}), we see that the
halo mass definition implicitly presumed in the excursion set
formalism is the SO($\Delta\vir$) one, with $\Delta\vir$ dependent on
time and cosmology. In the Einstein-de Sitter universe, where
$\delta\co(t)$ is equal to $3(12\pi)^{2/3}/20$ and $D(t)=a(t)$,
$\Delta\vir(t)$ takes the constant value $18\pi^2\approx 178$. While,
at $t_0$ in the concordant model, where $\delta\co(t_0)$ and $D(t_0)$
are respectively equal to $\approx 1.674$ and $\approx 0.760$,
$\Delta\vir(t_0)$ takes the value $\approx 359$
\citep{H00}\footnote{According to equation (\ref{Deltavir}),
  $\Delta\vir(t_0)=359$ and $D(t_0)=0.760$ imply
  $\delta\co(t_0)=1.621$ rather than 1.674. This 3.5\% error arises
  from the neglect of the cosmological constant in equation
  (\ref{Deltavir}).}.

\subsection{The CUSP Formalism}\label{JSDMcase}

In ellipsoidal collapse, the total energy $E\p$ of a sphere with mass
$M$ is not conserved. On the other hand, for peaks (hence, with outwards
decreasing density profiles) shells exchange energy as they cross each
other, causing virialisation to progress from the centre of the system
outwards. Thus, the spherical top-hat model does not hold. However, as
shown in SVMS, one can still accurately derive the typical density
profile $\rho\h(r)$ for haloes.

The variation in time of the total energy of a sphere undergoing
ellipsoidal collapse compared to that of the spherically averaged
system can be accurately monitored. In addition, during virialisation,
there is no apocentre-crossing (despite there being shell-crossing),
which causes virialised haloes to develop {\it from the inside out},
keeping the instantaneous inner structure unchanged. In these
conditions, the radius $r$ encompassing any given mass $M$ in the
final triaxial virialised system {\it exactly} satisfies the
relation\footnote{Again, the effects of the cosmological constant at
  halo scales are neglected.}
\beq 
r=-\frac{3GM^2}{10 E\p(M)},
\label{mr2bis}
\eeq
identical, at every radius $r$, to the relation (\ref{mr2}) holding
for the whole object in the excursion set case. 

In equation (\ref{mr2bis}), $E\p(M)$ is the (now non-conserved) energy
distribution of the spherically averaged protohalo. In the parametric
form, it is given by
\beq 
E\p(r)=4\pi\!\int_0^{r}\!\!\der \tilde r\, \tilde r^2
\rho\p(\tilde r)
\!\left\{\!\frac{\left[H_{\rm i}
    \tilde r-v\p(\tilde r)\right]^2}{2}\!-\!\frac{GM(\tilde r)}{\tilde r}
\!\right\}
\label{E1}
\eeq
\beq 
M(r)=4\pi\! \int_0^{r} \der \tilde r\, \tilde r^2\, \rho\p(\tilde r)\,,
\label{M1}
\eeq 
where $\rho\p(r)=\bar\rho\ii[1+\delta\p(r)]$ is the (unconvolved)
spherically averaged density profile of the protohalo, $H_{\rm i}$ is
the Hubble constant at $\ti$ and
\begin{equation}
  v_p(r)=\frac{2G\,\delta\!M(r)}{3H_{\rm i}r_p^2}
\label{vp}
\end{equation}
is, to leading order in the perturbation, the peculiar velocity at $r$
induced by the inner mass excess (e.g. \citealt{Peebles}),
\beq
\delta\!M(r)= 4\pi \int_0^{r} \der \tilde r\,\tilde r^2\,\bar\rho\ii\,\delta\p(\tilde r)\,.
\label{dM}
\eeq 
Replacing $v\p(r)$ given by equation
(\ref{vp})--(\ref{dM}) into equation (\ref{E1}), we are led to
\beq
E\p(r)=-\frac{20\pi}{3} \int_0^{r}\der \tilde r\, \tilde r\,
\rho\p(\tilde r)\, G\,\delta\!M(\tilde r)\,.
\label{E2}
\eeq
Note that the non-null (in Eulerian coordinates) peculiar velocity
$v\p(r)$ introduces a factor 5/3 in the value of $E\p(r)$ with respect
to the one resulting in the absence of peculiar
velocities. In the usual presentation
(in Lagrangian coordinates) of spherical collapse, $v\p(r)$ is
null, but the initial density contrast then decomposes in the
growing and decaying modes and the mass excess causing the
gravitational pull in equation (\ref{vp}) has an extra factor 5/3
compared to the mass excess $\delta\!M(r)$ associated with the growing
mode $\delta\p$ contributing to the mass of the final
halo. Consequently, the resulting value of $E\p(r)$ is exactly the
same as in equation (\ref{E2}).

The density contrast $\delta(R)$ of the seed of the progenitor of
scale $R$ of any accreting halo with $M$ at $t$ is but the
(unconvolved) spherically averaged density contrast profile
$\delta\p(r)$ of the protohalo convolved with a Gaussian window of
radius $R$. We thus have
\beq
\delta(R)=\frac{4\pi}{(2\pi)^{3/2}R^3}
\int_0^{\infty}\der r\, r^2\,\delta\p(r)\,{\rm e}^{-\frac{1}{2}\left(\frac{r}{R}\right)^2}\,.
\label{dp1}
\eeq 
Equation (\ref{dp1}) indicates that the trajectory $\delta(R)$ of
peaks with varying scale $R$ tracing the accretion of a halo with mass
$M$ at $t$ is the Laplace transform of the profile $\delta\p(r)$ of
its seed. As for haloes growing inside-out the mean density profile
$\rho\h(r)$ is determined by the mean accretion rate $\der M/\der t$
undergone over their aggregation history, the mean peak trajectory
$\delta(R)$ tracing such an evolution is characterised by having the
mean slope $\der R/\der \delta$ of peaks with $\delta$ at every $R$. Thus,
the desired mean peak trajectory $\delta(R)$ is the solution of the
differential equation\footnote{The distribution of peak curvatures $x$
  given in the Appendix A is a quite peaked symmetric function,
  so the inverse of the mean inverse curvature, $\lav x^{-1}\rav$,
  is close to the mean curvature, $\lav x\rav$.}
\beq 
\frac{\der \delta}{\der R}= - \,\lav x\rav[R,\delta(R)]\,\sigma_2(R)\,R\,
\label{dp2}
\eeq
for the boundary condition $\delta[R=\R(M,t)]=\delta\pk(t)$.  Once the
trajectory $\delta(R)$ has been obtained, we can infer the profile
$\delta\p(r)$ by inversion of equation (\ref{dp1}) (see SVMS for
details) and use equations (\ref{M1}) and (\ref{E2}) to calculate
$E\p(M)$. Then, replacing this function into equation (\ref{mr2bis}),
we can infer the mass profile $M(r)$ of the halo, leading to the
density profile $\rho\h(r)$, which turns out to be in good
agreement with the results of simulations (JSDM).

The mean spherically averaged halo density profile thus depends, like
the MF itself, on the particular mass definition adopted through the
functions $\delta\pk(t)$ and $\R(M,t)$ or, equivalently,
$\delta\pko(t)$ and $q(M,t)$, setting the boundary condition for
integration of equation (\ref{dp2}).

To obtain the halo mass definition that corresponds to any given pair
of $\delta\pko(t)$ and $q(M,t)$ functions, we must calculate the
density profile for haloes of different masses and find the relations
$f_1[\rho\h(R),M]=\bar\rho(t)$ and $f_2[\bar\rho\h(R),M]=\bar\rho(t)$
between $\rho\h(R)$ and $\bar\rho\h(R)$ and the mean cosmic density
$\bar\rho(t)$ for haloes with different $M$. Then, inverting, say, $f_2$
so as to obtain $M$ as a function of $\bar\rho\h(R)$ and $\bar\rho(t)$ and
replacing it in $f_1$, we are led to a relation
\beq
F[\rho\h(R),\bar\rho\h(R)]=\bar\rho(t) 
\eeq 
of the general form of relations (\ref{first}) and (\ref{second})
setting the halo mass definition associated with the CUSP MF with
arbitrary functions $\delta\pko(t)$ and $q(M,t)$.

Conversely, the functions $\delta\pko(t)$ and $q(M,t)$ can be inferred
from any given halo mass definition. To do this we must impose the
two following consistency arguments (JSDM): i) the total mass
associated with the resulting density profile must be equal to $M$ and
ii) the resulting MF must be correctly normalised.  For
SO($\Delta\vir$) or FoF(0.19) masses in the concordant cosmology, JSDM
found
\beq
\delta\pko(t)=\delta\co(t)\frac{[a(t)]^{1.0628}}{D(t)}\,
\label{deltas}
\eeq
and
\beq
q(M,t)\approx \left[Q\,\frac{\sigma_0\F(M,t)}{\sigma_0(M,t)}\right]^{-2/(n+3)}\,.
\label{qnu}
\eeq 
In equation (\ref{qnu}), the ratio $\sigma\F_0(M,t)/\sigma_0(M,t)$ takes the form
\beqa 
\frac{\sigma\F_0(M,t)}{\sigma_0(M,t)}=1- 0.0682\left[\frac{D(t)}{D(t_0)}\right]^{2}\nu\,,
\label{sigmas}
\eeqa
$Q$ is defined as 
\beq
Q^2\equiv  \frac{\int_0^\infty \der x \, x^{n+2}\,W_{\rm G}^2(x)}{\int_0^\infty \der x \, x^{n+2}\,W^2_{\rm TH}(x)}\,,
\label{A}
\eeq 
where $W_{\rm th}(x)$ and $W_{\rm G}(x)$ are the Fourier transforms of the
top-hat and Gaussian windows of radius $x/k$, respectively, and $n$ is
the effective spectral index. The approximate relation (\ref{qnu})
follows from the more fundamental one (\ref{sigmas}), taking into
account the relation
\beq
\sigma^2_0(\R)\approx\frac{A}{2\pi^2} \R^{-(n+3)} \int_0^\infty \der x\, W^2(x)\, x^{n+2}\,,
\label{j2}
\eeq
holding for the 0th order spectral moment for a filter with Fourier
transform $W$ under the power-law approximation, $P(k)=Ak^n$, of the
CDM (linear) spectrum, and the fact that the Gaussian and top-hat
radii for a seed with $M$ are respectively equal to $\R$ or to
$q\p\R$. This means that $Q$ and $n$ depend on the mass range
considered. For the mass ranges typically covered by the MFs found in
simulations, $Q$ and $n$ take values around $0.5$ and $-1.5$,
respectively.

Therefore, the CUSP MF is more general than the excursion set one in
the sense that it does not presume any particular mass definition; it
holds for any arbitrary one, adapting to it through the functions
$\delta\pko(t)$ and $q(M,t)$. 

\section{Similarity of SO and FoF Masses}\label{simil}

The fact that the CUSP formalism distinguishes between different mass
definitions can be used to try to understand the origin of the
similarity between SO and FoF masses and their respective mass and
multiplicity functions.

Equations (\ref{M1}) and (\ref{E2}) imply
\beq 
\frac{\der E\p}{\der M}\!=-\frac{5G\,\delta
  \!M(R\p)}{3R\p}\!=-\frac{5}{3}
\left[\frac{4\pi\bar\rho\ii}{3}\right]^{1/3}\!G
  M^{2/3}\delta\pk\F(M,t)\,,
\label{new}
\eeq 
where we have taken into account that the radius $R\p$ of the
protohalo is equal to $q\p\R$ with $q\p$ satisfying equation
(\ref{rmtru}). Comparing with the $M$-derivative of equation
(\ref{mr2bis}) and taking into account the identity
$M=4\pi\bar\rho\h(R\h)R\h^3/3$, equation (\ref{new}) leads to the
relation
\beq
\frac{5}{9}\left[\frac{\bar\rho\ii}{\bar\rho(t)}\right]^{1/3}\delta\pk\F(M,t)=\!\!\left[\frac{\bar\rho\h(R\h)}{\bar\rho(t)}\right]^{1/3}\!\left[1-\frac{\bar\rho\h(R\h)}{6\,\rho\h(R\h)}\right]\!,
\label{rho}
\eeq
which, making use of the definition of $F(c)$, can be rewritten in the
two following forms
\beq
\bar\rho\h(R)=\bar\rho(t)\!\left[\frac{5\,\delta\pk\F(M,t)a(t)}{9\,a(\ti)}\right]^{3}\left[1-\frac{F(c)}{6}\right]^{-3}
\label{rho2}
\eeq
and
\beq
\rho\h(R)=\bar\rho(t)\!\left[\frac{5\,\delta\pk\F(M,t)a(t)}{9\,a(\ti)]}\right]^{3}\left[1-\frac{F(c)}{6}\right]^{-3} \frac{1}{F(c)}\,.
\label{rho3}
\eeq
For SO and FoF masses, these expressions therefore imply
\beq
\Delta=\left[\frac{5\,\delta\pk\F(M,t)a(t)}{9\,a(\ti)}\right]^3\left[1-\frac{F(c)}{6}\right]^{-3}
\label{Delta2}
\eeq
and
\beq
b=\left[\frac{2\pi}{3F(c)}\right]^{-1/3}\left[\frac{5\,\delta\pk\F(M,t)a(t)}{9\,a(\ti)}\right]^{-1}\left[1-\frac{F(c)}{6}\right]\,,
\label{b}
\eeq
respectively. 

Equation (\ref{Delta2}) seems to indicate that, in the SO case, {\it
  the mass dependence} of $\delta\pk\F$ must cancel with that coming
from $F(c)$. But equation (\ref{first}) implies $R\h\propto M^{1/3}$,
which, replaced into equation (\ref{mr2bis}) at $r=R\h$, leads to
$E\p(M)\propto M^{5/3}$ and, hence, to $\der E\p/\der M\propto
M^{2/3}$, implying (see eq.~[\ref{new}]) that $\delta\pk\F$ is {\it a
  function of $t$ alone}. The solution to this paradox is that, to
leading order in the perturbation as used in the derivation of the
density profile (see eq.~[\ref{vp}]), $\delta\pk\F$ and $F(c)$ are, in
the SO case, independent of $M$. (Likewise, eq.~[\ref{b}] multiplied
by the cubic root of $F(c)$ leads in the FoF case to a similar
paradox, with identical solution.) Consequently, to such an order of
approximation, the SO and FoF mass definitions with $\Delta$ and $b$
satisfying equation (\ref{cor}) are equivalent to each other.

We thus see that the origin of this approx equivalence is the
inside-out growth of accreting haloes, crucial to obtain equation
(\ref{mr2bis}) setting the typical spherically averaged density
profile for haloes arising from peaks that undergo ellipsoidal
collapse and virialisation. But this is not all. We can go a step
further and infer the value of $b$ leading to FoF masses 
equivalent to SO($\Delta\vir$) ones. 

The relation between the two functions (of $t$) $\delta\pk\F$ and
$1-F(c)/6$ can be readily derived for the particular case of
SO($\Delta\vir$) haloes. Comparing equations (\ref{mr2}) and
(\ref{mr2bis}), the latter at $r=R\h$, we have that haloes arising
from ellipsoidal collapse of peaks with $\delta\pk$ in the density
field at $\ti$ smoothed with a Gaussian filter of radius $\R$, could
have formed according to the spherical top-hat model from the same
seeds with $\delta\pk\F$ when the density field is smoothed with a
top-hat filter of radius $q\R$.\footnote{The outwards decreasing
  density profile of seeds for {\it purely accreting haloes} ensures
  the possibility to use any spherical window to define the one-to-one
  correspondence between haloes and peaks. The use of a Gaussian
  window is only mandatory, as mentioned, if haloes can also 
  undergo major mergers (see MSS and SVMS).}  Equations
(\ref{Deltavir}) and (\ref{Delta2}), the latter for
$\Delta=\Delta\vir$, then imply
\beq 
\delta\pk\F(t)=\delta\cc(t)\,
  6\left[1-\frac{F(c)}{6}\right]\,.
\label{thdeltas}
\eeq
The typical value of $F(c)$ for SO($\Delta\vir$) haloes can be
inferred from equation (\ref{thdeltas}) for $\delta\cc(t)$ given by
equation (\ref{delta1}) and $\delta\pk\F$ given by equation
(\ref{delta}) for seeds of any arbitrary mass. However, the density
profile $\delta\p(r)$ of protohaloes is not accurate enough (owing to
the inverse Laplace transform of eq.~[\ref{dp1}]) for $\delta\pk\F$ to
be inferred with the required precision. Therefore, as the CUSP
formalism recovers, to leading order in the perturbation, the typical
spherically averaged density profile for simulated haloes, we can
estimate $F(c)$ directly from such empirical profiles. As well-known
these profiles are of the NFW form \citep{NFW97} and, hence, satisfy
the relation
\beq 
F(c)\equiv \frac{\bar\rho\h(R\h)}{\rho\h(R\h)}=
3\,\frac{(1+c)^2}{c^2}\left[\ln(1+c)-\frac{c}{1+c}\right]\,.
\label{Fc}
\eeq
For $c$ spanning from $\sim 5$ to $\sim 15$ as found in simulations of
the concordant cosmology for SO($\Delta\vir$) haloes at $t_0$ (and
approximately at any other time and cosmology), we find $F(c)\sim 5.1
\pm 0.1$. And, bringing this value of $F(c)$ and
$\Delta=\Delta\vir\approx 359$ into equation (\ref{cor}), we arrive at
$b\sim 0.19$, in full agreement with the results of numerical
simulations. Of course, the exact typical value of $F(c)$ may vary
with time and cosmology. But, according to the results of numerical
simulations, we do not expect any substantial variation in this sense,
so we have that FoF(0.2) masses are approximately equivalent to
SO($\Delta\vir$) ones, in general.

As a byproduct we have that equation (\ref{thdeltas}) for $F(c)\approx
5.1 \pm 0.1$ implies the relation $\delta\pko\F(t)\sim 0.9\,
\delta\co(t)$. In other words, in the case of SO($\Delta\vir$) or
FoF(0.2) masses, the top-hat density contrast for ellipsoidal collapse
and virialisation would take an almost universal value independent of
$M$, just a little smaller than the almost universal value
$\delta\co(t)$ for spherical collapse. This result thus suggests that
it should be possible to modify the excursion set formalism in order
to account for ellipsoidal collapse and virialisation by simply
decreasing the usual density contrast for spherical collapse by a
factor $\sim 0.9$. We will comeback to this interesting prediction
below.

\section{Multiplicity Function}\label{univ}

The multiplicity function associated with any given MF, $\partial
n(M,t)/\partial M$, is defined as
\beq 
f(\sigma_0\F,t)=\frac{M}{\bar\rho}\frac{\partial n[M(\sigma_0\F),t]}{\partial \ln
  [(\sigma_0\F)^{-1}]}\,.
\label{f}
\eeq

In the excursion set case, this leads to a function of the simple
form
\beq f_{\rm es}(\sigma_0\F,\delta\cz)=\left(\frac{2}{\pi}\right)^{1/2}\,\nu\ES\,{\rm
  e}^{-\frac{\nu\ES^2}{2}}\,,
\label{PSmf}
\eeq
while, in the CUSP case, it leads to (see eqs.~[\ref{mf}] and
[\ref{f}])
\beq f_{\rm
  CUSP}(\sigma_0,\delta\pkz)=\frac{M(\sigma_0,\delta\pkz)}{\bar\rho}\,
N\nest(\sigma_0,\delta\pkz)\,.
\label{CUSPmf}
\eeq
To obtain equation (\ref{CUSPmf}) we have taken the partial derivative
of $n_{\rm CUSP}$ with respect to $\sigma_0$ instead of $\sigma_o\F$
as prescribed in equation (\ref{f}). But this is irrelevant for
SO($\Delta\vir$) or FoF(0.19) masses in the concordant cosmology as
hereafter assumed, given the relation (\ref{sigmas}) between the two
0th order spectral moments.

\subsection{Comparison with Simulations}

\begin{figure}
\centerline{\includegraphics[scale=0.46]{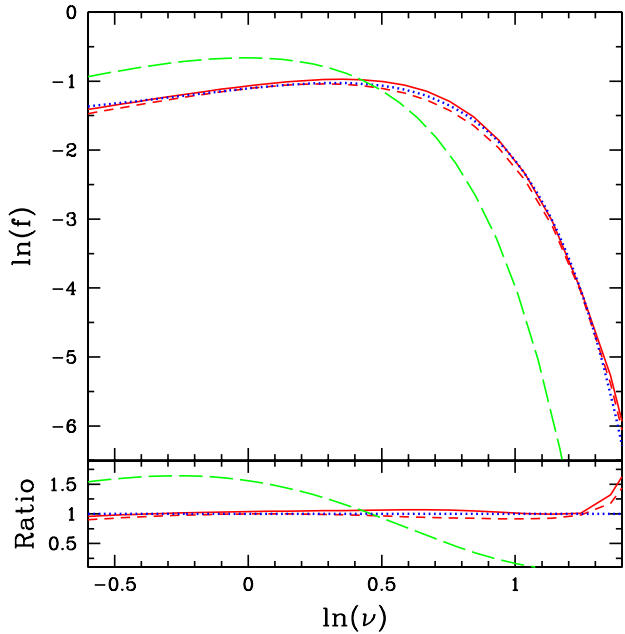}}
\caption{Multiplicity function at $t_0$ derived from the CUSP (red
  lines) and excursion set (green long-dashed line) formalisms,
  compared to \citet{Wea06} analytic fit to the multiplicity function
  of simulated haloes (blue dotted line) over the maximum mass range
  $(2\times 10^{10}$ \modotc, $2\times 10^{15}$ \modotc) covered by
  simulations. For the CUSP case, we plot both the the approximate
  solution not accounting for the correlation between peaks of
  different scales (dashed line) and the more accurate solution given
  in Appendix (solid line). Ratios in the bottom panel are with
  respect to $f_{\rm W}$.}
\label{f1}
\end{figure}
\begin{figure}
\centerline{\includegraphics[scale=0.46]{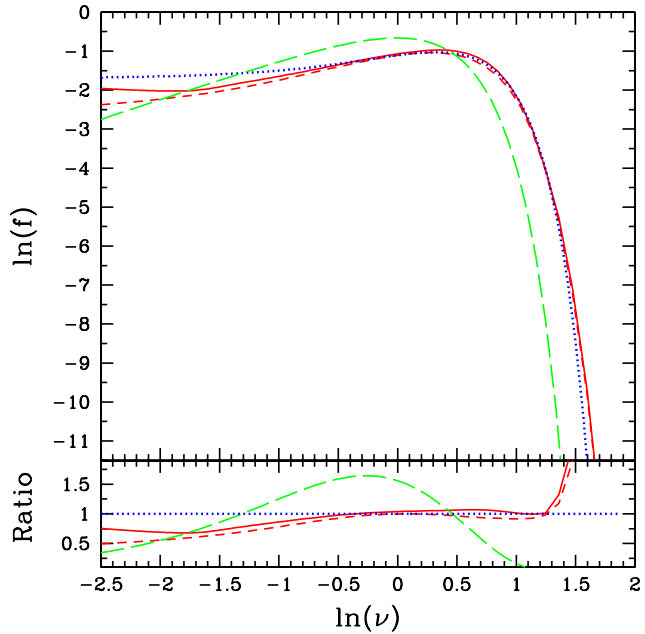}}
\caption{Same as Figure \ref{f1} but in a much wider mass range,
  corresponding to current haloes with masses from $1$ \modot to
  $3\times 10^{16}$ \modotc.}
\label{f2}
\end{figure}

In Figure \ref{f1} we compare these two multiplicity functions at
$t_0$ to the \citet{Wea06} analytic expression, of the \citet{ST02}
form,
\beq
f_{\rm W}(\nu_{\rm es})=0.3303\left(\nu_{\rm es}^{1.625}+0.5558\right)\,{\rm e}^{-0.4565\nu_{\rm es}^2}\,,
\label{Wmf}
\eeq
fitting the multiplicity function of simulated haloes with FoF(0.2)
masses at $t_0$ in all CDM cosmologies. $f_{\rm W}$ is usually
expressed as a function of $\sigma_0\F$ instead of $\nu_{\rm es}$; the
expression (\ref{Wmf}) has been obtained from that usual expression assuming
$\delta\co(t_0)=1.674$ (taking the value 1.686 would make no significant
difference). In Figure \ref{f1}, all the multiplicity
functions are expressed as functions of the Gaussian height for
ellipsoidal collapse and virialisation, $\nu$, instead of the top-hat
height for spherical collapse, $\nu_{\rm es}$. The change of variable
from $\nu\ES$ to $\nu$ has been carried out using the relation
\beq
\nu\ES=\frac{D(t)}{[a(t)]^{1.0628}}\,\nu\left\{1-0.0682\left[\frac{D(t)}{D(t_0)}\right]^{2}\nu\right\}^{-1}
\label{nus2}
\eeq
that follows from equations (\ref{deltas}) and (\ref{sigmas}). This is
a mere change of variable; it does not presume any modification in the
assumptions entering the derivation of the different multiplicity
functions.

As can be seen, while $f_{\rm es}$ shows significant deviations from
$f_{\rm W}$ at both mass ends, $f_{\rm CUSP}$ is in excellent
agreement with $f_{\rm W}$ all over the mass range covered by
simulations. This is true regardless of whether we consider the
approximate or more accurate versions of $f_{\rm CUSP}$. The deviation
(of opposite sign in both cases) is less than 6.5\%. We stress that
there is no free parameter in the CUSP formalism, so this agreement is
really remarkable.

It might be argued that $f_{\rm CUSP}$ cannot be trusted at small
$\nu$'s because peaks with those heights have big chances to be
destroyed by the gravitational tides of neighbouring massive
peaks. Although this possibility exists, peaks suffering strong tides
are expected to be nested within such neighbours and,
hence, they should not be counted in the MF corrected for
nesting. The correction for nesting becomes increasingly important, indeed,
towards the small $\nu$ end. On the other hand, $f_{\rm CUSP}$ is
well-normalised\footnote{The CUSP MF is well-normalised by
  construction as this is one of the conditions imposed to obtain the
  functions $\delta\pko(t)$ and $q(M,t)$.} and still predicts the right
abundance of massive haloes, which would hardly be the case if $f_{\rm
  CUSP}$ overestimated the abundance of low-mass objects. Therefore,
we do not actually expect any major effect of that kind.

It is thus worth seeing how $f_{\rm CUSP}$ compares to $f_{\rm W}$
outside the mass range covered by simulations. In Figure \ref{f2} we
represent the same multiplicity functions as in Figure \ref{f1} over a
much wider range. Surprisingly, the agreement between $f_{\rm CUSP}$
and $f_{\rm W}$ is still very good. At very small $\nu$'s, $f_{\rm W}$
shows a slight trend to underestimate the abundance of haloes
predicted by $f_{\rm CUSP}$, but the difference is small. It increases
monotonously until reaching, in the case of the accurate version of
$f_{\rm CUSP}$, a ratio of $\sim 0.70$ ($\sim 30$\% deviation) at
$M\sim 5\times 10^4$ \modotc.

\subsection{Approx Universality}

The excursion set multiplicity function expressed as a function of
$\nu\ES$, $f_{\rm es}(\nu_{\rm es})$, is cosmology-independent (it
takes the same form [\ref{PSmf}] in all cosmologies) and
time-invariant (the height is constant). Hence, it is universal in a
strict sense. Such a universality is in fact what has motivated the use of
the multiplicity function defined in equation (\ref{f}) instead of the
(non-universal) MF. Unfortunately, $f_{\rm es}$ does not properly
recover the multiplicity function of simulated haloes.

But $f_{\rm CUSP}$ does, so the question rises: is $f_{\rm CUSP}$ also
universal?  Certainly, since the CUSP MF (as well as the real MF of
simulated haloes) depends on the particular halo mass definition while
$\sigma_0\F$ does not, $f_{\rm CUSP}$ will necessarily depend (like
the multiplicity function of simulated haloes; see
e.g. \citealt{Tea08}) on the mass definition adopted. Thus, we will
focus on the SO($\Delta\vir$) or FoF(0.2) mass definitions, as
suggested by the results of simulations (see the form [\ref{Wmf}] of
$f_{\rm W}(\nu_{\rm es})$).

\begin{figure}
\centerline{\includegraphics[scale=0.46]{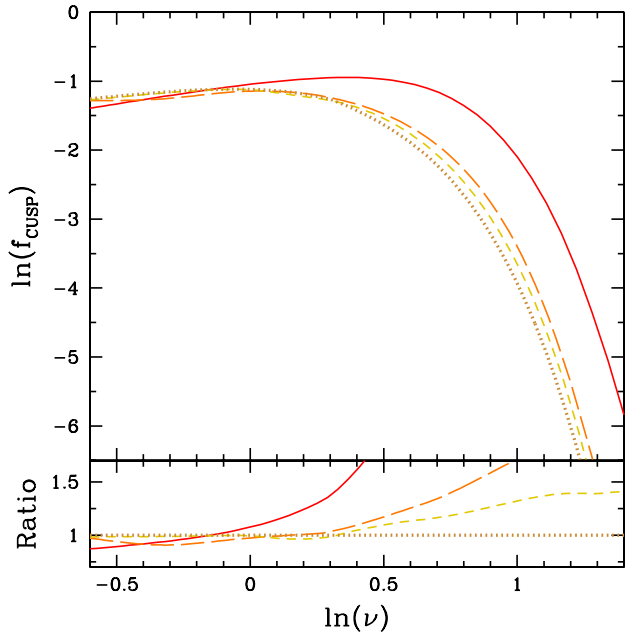}}
\caption{CUSP multiplicity functions at $z=0$, 5, 10 and 20, from left
  to right, in red (solid line), orange (long-dashed line), gold
  (dashed line) and brown (dotted line),
  respectively. Ratios in the bottom panel are with respect to the
  multiplicity function at $z=20$.}
\label{f3}
\end{figure}
\begin{figure}
\centerline{\includegraphics[scale=0.46]{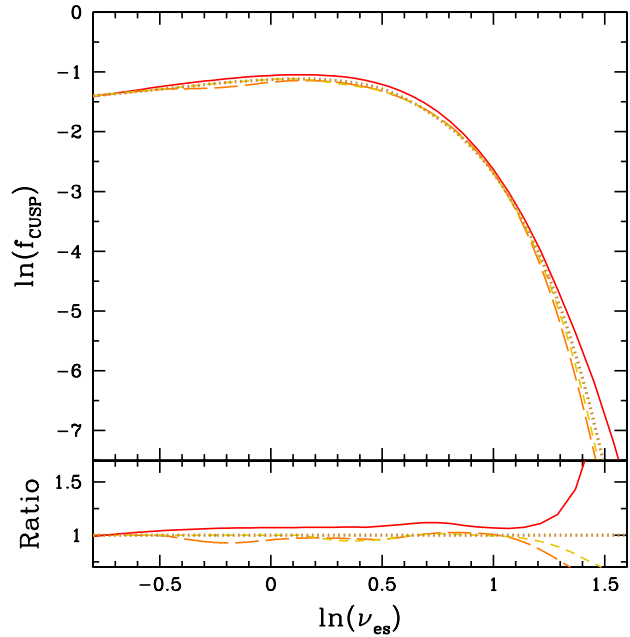}}
\caption{Same as Figure \ref{f3} but for $f_{\rm CUSP}$ expressed as a
  function of variable $\nu\ES$ instead of $\nu$. The green
  long-dashed line represents the multiplicity function that would be
  obtained from the excursion set formalism taking a density contrast
  for spherical collapse equal to 0.89 times the usual value.}
\label{f4}
\end{figure}

By construction, the unconditioned and conditional peak number
densities, $N(\sigma_0,\delta\pko)$ and
$N(\sigma_0,\delta\pko|\sigma_0',\delta\pko)$ entering the Volterra
equation (\ref{nnp}) take the same form of $\sigma_0$, $\sigma'_0$ and
$\delta\pkz$, through the heights $\nu$ and $\nu'$, in all cosmologies
(see eqs.~[\ref{npeak}], [\ref{int2}] and
[\ref{averx}])\footnote{$\epsilon$ takes the form
  $2^{(n+3)/2}(\nu/\nu')[1+(\nu/\nu')^{4/(n+3)}]^{-(n+3)/2}$, where
  $n$ is the effective spectral index in the relevant mass
  range.}. Certainly, these number densities also depend on $\gamma$,
$\gamma'$ and $R_\star$ that involve spectral moments of different
orders and, hence, depend on the cosmology through the exact shape of
the (linear) power spectrum. However, in all CDM cosmologies, the
effective spectral index $n\eff$ takes essentially the same fixed
value, with less than 20\% error over the whole mass range $(2\times
10^{10}$ \modotc, $2\times 10^{15}$ \modotc) of interest, implying
that $\gamma\approx \gamma'$ and $R_\star/\R [3(1-\gamma^2)]^{1/2}$
takes almost ``universal'' values respectively equal to $0.6\pm 0.1$
and $1.4\pm 0.1$. Thus, those number densities are indeed very
approximately universal functions of $\nu$ and $\nu'$ {\it but for a
  factor $\R^{-3}$}. Moreover, if we multiply the Volterra equation
(\ref{nnp}) by $M/\bar\rho=4\pi \bar\rho\ii(q\R)^3/(3\bar\rho)$ so
that its solution is directly $f_{\rm CUSP}$ (see eq.~[\ref{CUSPmf}]),
then the factor $\R^{-3}$ in the two number densities cancels with the
factor $\R^3$ coming from the mass. Therefore, the solution $f_{\rm
  CUSP}$ of such a Volterra equation will have very
approximately the same expression of $\nu$ in all CDM cosmologies,
provided only the function $q(M,t)$ does.

But, according to equations (\ref{qnu})--(\ref{sigmas}) holding for
SO($\Delta\vir$) and FoF(0.19) haloes, $q(M,t)$ involves the ratio
$\sigma\F_0/\sigma_0$ which is not a function of $\nu$ alone, but also
depends on $t$ through the cosmology-dependent ratio
$D(t)/D(t_0)$. Nevertheless, the term with the ratio
$\sigma\F_0/\sigma_0$ responsible of the undesired functionality of
$q(M,t)$ is small in general (except for large $\nu$'s), particularly
at high-$z$ where $D(t)/D(t_0)$ becomes increasingly small. There,
$q(M,t)$ becomes constant (equal to $Q^{-2/(n+3)}$) and $f_{\rm
  CUSP}(\nu)$ becomes essentially universal. However, at low-$z$ this
is only true for small enough $\nu$'s.

In Figure \ref{f3} we show $f_{\rm CUSP}(\nu)$ in the concordant model
for various redshifts (see JSDM for the corresponding MFs, in full
agreement with the results of simulations). The deviations from
universality or, more exactly, from time-invariance at high-$z$ are
small as expected, but at low-$z$ they are very marked. Thus, $f_{\rm
  CUSP}(\nu)$ is far from universal (!).

But this result was not unexpected. Given the relation (\ref{nus2})
between $\nu$ and $\nu_{\rm es}$, we cannot pretend that $f_{\rm
  CUSP}(\nu)$ is universal as a function of both arguments at the same
time. Inspired by the universality of $f_{\rm es}(\nu_{\rm es})$, most
efforts in the literature have been done in trying to find one mass
definition rendering the multiplicity function of simulated haloes
approximately universal {\it as a function of the top-height for
  spherical collapse}, not as a function of the (unknown) Gaussian
height for ellipsoidal collapse and virialisation. Therefore, what we
should actually check is whether $f_{\rm CUSP}$ is universal as a
function of $\nu\ES$ and not of $\nu$. As shown in Figure \ref{f4},
when the change of variable from $\nu$ to $\nu\ES$ is made, $f_{\rm
  CUSP}$ becomes indeed almost fully time-invariant. Strictly, it
still shows slight deviations from universality at large $\nu\ES$, but
these deviations are in full agreement with those found in simulations
(see Fig.~14 in \citealt{Lea07}).

Thanks to the CUSP formalism, we can determine the Gaussian height for
ellipsoidal collapse and virialisation corresponding to any desired
mass definition. Thus, we can seek the halo mass definition for which
$f_{\rm CUSP}$ expressed as a function of $\nu$ takes a universal
form. According to the reasoning above, for this to be possible the
ratio $\sigma_0\F/\sigma_0$ should be equal to $1+c\, \nu$, with $c$
equal to an arbitrary universal constant. This would ensure both that
the partial derivative of $n_{\rm CUSP}$ with respect to $\sigma_0$
coincides with the partial derivative with respect to $\sigma_0\F$ and
that the function $q(M,t)$ is a function of $\nu$ alone:
$q(\nu)\approx [Q\,\left(1+c\,\nu\right)]^{-2/(n+3)}$.  Consequently,
following the procedure given in Section \ref{JSDMcase}, we can infer,
from such a function $q(\nu)$ and any arbitrary function
$\delta\pko(t)$, the desired halo mass definition. Unfortunately,
despite the freedom left in those two functions, the mass definition
so obtained will hardly coincide with any of the practical SO and FoF
ones. Thus, it is actually preferable to keep on requiring the universality of
the multiplicity function in terms of $\nu\ES$ as usual.

But this does not explain why the FoF mass definition with linking
length $\sim 0.2$ is successful in giving rise to a universal
multiplicity function expressed as a function of $\nu\ES$.  Clearly,
what makes this mass definition special is that, for the reasons
explained in Section \ref{simil}, it coincides with the
SO($\Delta\vir$) definition. In fact, as mentioned there, the exact
value of the linking length may somewhat vary with time and cosmology,
so the canonical mass definition would be the SO($\Delta\vir$)
definition rather than the FoF(0.2) one. But why should a mass
definition that involves the virial overdensity $\Delta\vir$ arising
from the formal {\it spherical top-hat model} successfully lead to a
universal multiplicity function expressed as a function of $\nu\ES$ if
haloes actually form from peaks that undergo {\it ellipsoidal collapse
  and virialisation}? The reason for this is that, as a consequence of
the inside-out growth of haloes formed from ellipsoidal collapse and
virialisation, they satisfy the relation (\ref{mr2bis}), identical to
the relation (\ref{mr2}) satisfied by objects formed in the spherical
top-hat model.

\begin{figure}
\centerline{\includegraphics[scale=0.46]{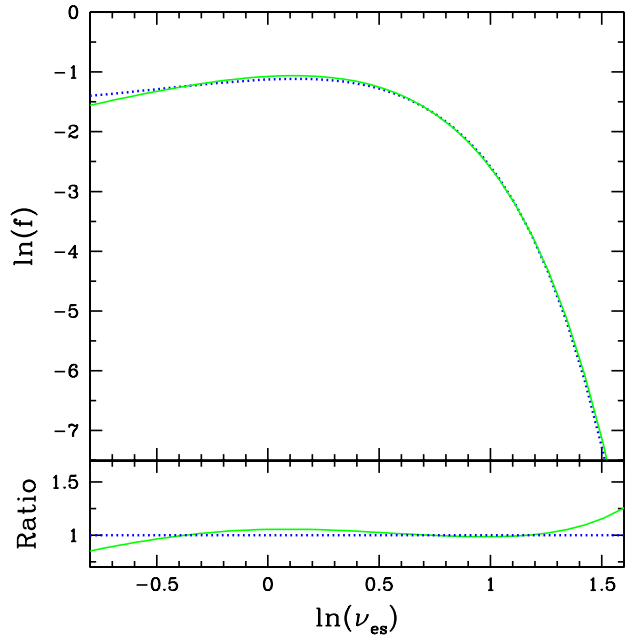}}
\caption{Modified excursion set multiplicity function resulting from a
  density contrast for collapse equal to 0.889 times the usual value
  (solid green line), compared to $f_{\rm W}$ for FoF(0.2) haloes
  (dotted blue line). Both multiplicity functions are strictly
  universal, so the two curves hold for any arbitrary redshift.}
\label{f5}
\end{figure}

As mentioned, an interesting consequence of this ``coincidence'' is
that the top-hat density contrast for ellipsoidal collapse and
virialisation for SO($\Delta\vir$) masses, $\delta\pko\F$, takes a
universal value, independent of $M$, approximately equal to 0.9 times
the top-hat density contrast for spherical collapse,
$\delta\co(t)$. Given this relation, changing the latter density
contrast by the former in the excursion set formalism, $f\ES(\nu\ES)$
should keep on being universal and, in addition, recover the real
multiplicity function of haloes formed by ellipsoidal collapse and
virialisation. As shown in Figure \ref{f5}, this is fully
confirmed. One must just renormalise the resulting modified excursion
set multiplicity function in the relevant mass range by multiplying it
by 0.714. But this is simply due to the fact that the correction for
nesting achieved in the excursion set formalism is inconsistent with
top-hat smoothing, which yields an increasing deviation of the
predicted function at low-masses, the most affected by such a
correction. (The right normalisation should naturally result if we
could implement the excursion set correction for nesting with top-hat
smoothing.)

Therefore, the ultimate reason for the success of the SO($\Delta\vir$)
mass definition, and by extension of the FoF(0.2) one, is that, as a
consequence of the inside-out-growth of accreting haloes, the
corresponding top-hat density contrast for ellipsoidal collapse and
virialisation is essentially proportional to the formal top-hat
density contrast for spherical collapse.

\begin{figure}
\centerline{\includegraphics[scale=0.46]{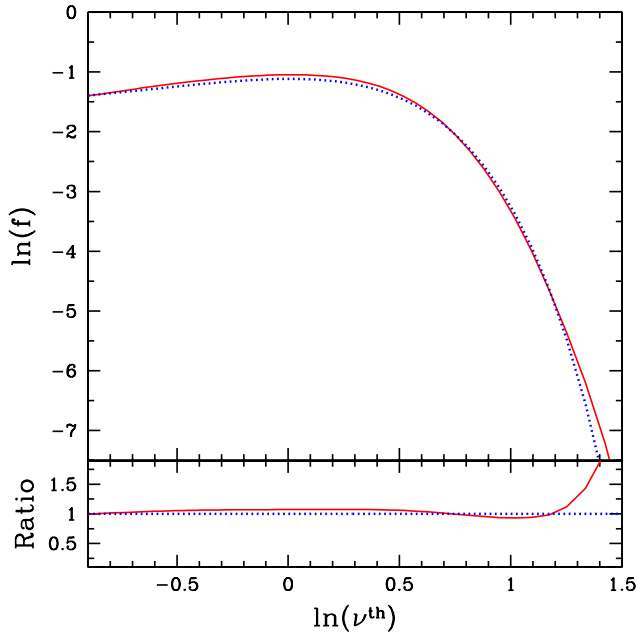}}
\caption{Same as Figures \ref{f3} and \ref{f4} but for $f_{\rm CUSP}$
  expressed as a function of the top-hat height for ellipsoidal
  collapse and virialisation, $\nu\F$, as suggested by the CUSP
  formalism. This multiplicity function would be strictly universal,
  so the curves for the different redshifts fully overlap. For
  comparison we also plot the function $f_{\rm W}$ (blue dotted curve)
  expressed with the same argument.}
\label{f6}
\end{figure}

To end up we want to mention that the previous result suggests what is
actually the most natural argument for the halo multiplicity function
to take a universal form: {\it the top-hat height for ellipsoidal
  collapse and virialisation}, $\nu\F$, defined as
$\delta\pkz\F(t_0)/\sigma\F_0\approx 0.889\,
\delta\cz(t_0)/\sigma\F_0$. This expression holds for the current time
and the concordant cosmology. The exact dependence of
$\delta\pko\F(t)$ (or, more exactly, of the ratio
$\delta\pko\F(t)/\delta\co(t)$) on time and cosmology is hard to tell
owing to the insufficient precision of the inverse Laplace transform
of equation (\ref{dp1}) or, alternatively, the unknown range of $c$
values of simulated haloes with SO($\Delta\vir$) masses at other times
and cosmologies. But a reasonable guess is that such a dependence
should make $f_{\rm CUSP}(\nu\F)$ be strictly universal and equal to
the multiplicity function represented in Figure \ref{f6}. The reason
for this guess is the full consistency, at any time and cosmology,
between the SO($\Delta\vir$) mass definition and the real dynamics of
collapse and virialisation of halo seeds. A similar full consistency
is what causes $f\ES(\nu\ES)$ to be also strictly universal. The
difference between the two cases is that, while the excursion set
formalism assumes a non-realistic dynamics of collapse (unless it is
modified as prescribed above), the CUSP formalism assumes the right
dynamics.

\section{Discussion and Conclusions}\label{discuss}

To have a fully satisfactory understanding of the role of the halo
mass definition in the properties of the halo multiplicity function we
must still answer one last question: why is the simplifying assumption
that ``haloes form by pure accretion'' used in all the derivations of
the halo MF so successful? As discussed in SVMS, major mergers go
unnoticed, indeed, in the typical spherically averaged density profile
and abundance of virialised haloes because the virialisation taking
place in such dramatic events is a real relaxation causing the memory
loss of the halo past history. In other words, virialised haloes do
not know whether they have suffered major mergers or they have formed
by pure accretion. As a consequence, to derive the halo MF (as well as
any other typical halo property, except those arising from two-body
relaxation) one has the right to assume, with no loss of generality,
that all haloes form by pure accretion. Note that, on the contrary,
the virialisation produced in smooth accretion is not a full
relaxation: the absence of apocentre-crossing during such a
virialisation (SVMS) preserves some memory of the initial seed. This
is the reason why the density profile for individual haloes can be
inferred from the respective peak trajectories $\delta(R)$ or,
equivalently, from their accretion history. The fact that the density
profile of accreting haloes remembers their mass aggregation history is
at the base of the well-known assembly bias.

JSDM showed that the typical spherically averaged density profile and
MF of haloes derived in the framework of the CUSP formalism recover
the results of numerical simulations for haloes for any chosen mass
definition. In the present paper, we have seen that the
SO($\Delta\vir$) and FoF(0.2) mass definitions are essentially
equivalent to each other and their respective multiplicity functions
expressed as a function of the top-hat height for spherical collapse
are very similar and approximately universal. The reason for those
trends is that, as a consequence of the inside-out growth of accreting
haloes, the top-hat density contrast for ellipsoidal collapse for
those particular mass definitions is essentially universal and equal
to $\sim 0.9$ times the usual top-hat density contrast for spherical
collapse.

\vspace{0.90cm} \par\noindent
{\bf ACKNOWLEDGEMENTS} \vspace{0.25cm}\par

\noindent This work was supported by the Spanish DGES
AYA2009-12792-C03-01 and AYA2012-39168-C03-02 and the Catalan DIUE 2009SGR00217.
One of us, EJ, was beneficiary of the grant BES-2010-035483.


\begin{appendix}

\section{Accurate Conditional Peak Number Density}\label{exact}

As shown in \citet{MSS98}, the conditional number density
$N\nest(\sigma_0,\delta\pko|\sigma'_0,\delta\pko)$ of peaks with
$\delta\pko$ per infinitesimal $\ln \sigma_0^{-1}$ subject to being
located in the collapsing cloud of {\it non-nested} peaks with
$\delta\pko$ at $\sigma'_0<\sigma_0$ is well-approximated by the
integral over the distance $r$ from the background peak out to the
radius $R\p$ of the collapsing cloud in units of $q(M,t)\R$ of the
conditional number density of peaks with $\delta\pko$ per infinitesimal
$\ln \sigma_0^{-1}$, subject to being located at a distance $r$ from a
background peak, $N(\sigma_0,\delta\pko|\sigma'_0,\delta\pko,r)$,
\beqa N\nest(\sigma_0,\delta\pko|\sigma'_0,\delta\pko)=
\!C\!\!\int_0^{1}\!\!
\der r\, 3 r^2 N(\sigma_0,\delta\pko|\sigma_0,\delta\pko,r)\,.
\label{int}
\eeqa
The conditional number density in the integrant on the right of
equation (\ref{int}) can be obtained, as the ordinary number density
(\ref{npeak}), from the conditional density of peaks per infinitesimal
$x$ and $\nu$, subject to being located at the distance $r$ from a
background peak with $\nu$ at $\sigma'_0$, calculated by BBKS. The
result is \citep{MSS98}
\beqa 
\,N(\sigma_0,\delta\pko|\sigma'_0,\delta\pko,r)\der \ln \sigma_0^{-1}~~~~~~~~~~~~~~~~~~~~~~\nonumber\\ =\frac{\lav
  x\rav[\tilde\sigma_0(r),\delta\pko]}{(2\pi)^2\,R_\star^3\,\gamma\,e(r)} {\rm
  e}^{-\frac{\left[\nu -\epsilon(r)\,\nu'(r)\right]^2}{2e^2(r)}}
  \der \ln \sigma_0^{-1}\,,
\label{int2bis}
\eeqa 
where $\lav x\rav[\tilde\sigma_0(r),\delta\pko]$ is the average
curvature of peaks with $\delta\pko$ at $\sigma_0$ located at a
distance $r$ from a background peak with identical density contrast at
$\sigma'_0$. This latter function takes just the same form as the
usual average curvature $\lav x\rav(\sigma_0,\delta\pko)$ for the
properly normalised (by integration over $x$ from zero to infinity)
curvature distribution function
\beq
h(x,\sigma_0,\delta\pko)=f(x)\,{\rm e}^{-{(x-x_\star)^2 \over 2(1-\tilde\gamma^2)}},
\label{mf17bis}
\eeq
\beqa
f(x)=\frac{x^3-3x}{2}\left\{{\rm erf}\!\left[\left(\frac{5}{2}\right)^{1/2}x\right]+{\rm erf}\!\left[\left(\frac{5}{2}\right)^{1/2}\frac{x}{2}\right]\right\}\nonumber\\
+\left(\frac{2}{5\pi}\right)^{\!\!1/2}\!\!\left[\left(\!\frac{31x^2}{4}+\frac{8}{5}\!\right){\rm e}^{-\frac{5x^2}{8}}+\left(\!\frac{x^2}{2}-\frac{8}{5}\!\right){\rm e}^{-\frac{5x^2}{2}}\right]\!,
\label{fx}
\eeqa
but for $\tilde x_\star(r)\equiv \tilde\gamma(r)\,\tilde\nu(r)$ instead of
$x_\star \equiv \gamma\,\nu$, being
\beq 
\tilde\gamma^2(r) =
\gamma^2\left[1 + \epsilon(r)^2\,{(1 - r_1)^2 \over 1 -
    \epsilon(r)^2}\right]
\label{tgamma}
\eeq
\beq
\tilde\nu(r)\!=\!{\gamma \over \tilde\gamma(r)}\,{1 - r_1 \over 1 -
\epsilon(r)^2}\left[\nu\!\left({1 - \epsilon(r)^2r_1 \over 1 - r_1}\!
\right)\!-\epsilon(r)\nu'(r)\right]\!.
\label{tildbis}
\eeq
In equations (\ref{int2bis}), (\ref{tgamma}) and (\ref{tildbis}), we
have used the following notation: $e(r)=\sqrt{1 - \epsilon(r)^2}$,
$\epsilon(r)=(\sigma_0^2(R\h)$ $/[\sigma_0\sigma'_0]g(r,\sigma'_0)$, and
$\nu'(r)=g(r,\sigma'_0)\overline{\delta(r)}/\sigma'_0$ and
$r_1=[\sigma_0(\R)\sigma_1(R\h) /(\sigma_1(\R)\sigma_0(R\h))]^2$, where
$R\h$ is defined as usual and $g(r,\sigma'_0)$ is
$\left\{1-[\Delta\delta'(r)]^2/\sigma'_0\right\}^{1/2}$, being
$\overline{\delta'(r)}$ and $\Delta\delta'(r)$ the mean and rms
density contrasts at $r$ from the background peak, respectively given
by
\beqa
\overline{\delta(r)}=\frac{\gamma\delta\pk}{1-\gamma^2}\left(\frac{\psi}{\gamma}+\frac{\nabla^2\psi}{u^2}\right)-\frac{x\sigma_0}{1-\gamma^2}\left(\gamma\psi+
\frac{\nabla^2\psi}{u^2}\right)
\label{e1bis}
\eeqa
\beqa
[\Delta\delta(r)]^2=\sigma_0^2\bigg\{\!1\!-\!\frac{1}{1-\gamma^2}\left[\psi^2+\!\!\left(2\gamma\psi+\frac{\nabla^2\psi}{u^2}\!\!\right)\frac{\nabla^2\psi}{u^2}\right]\nonumber\\
-5\left(\frac{3\psi'}{u^2r}-\frac{\nabla^2\psi}{u^2}\right)^2-\frac{3(\psi')^2}{\gamma u^2}\bigg\}\,,~~~~~~~~~~~~~~~~~~~~
\label{e2bis}
\eeqa
where $\xi(r)$ is the mass correlation function at the separation $r$
and scale $\R$, $\psi$ is the ratio $\xi(r)/\xi(0)$, $\psi'$ is its
$r$-derivative and $u$ is defined as
$[q(M,t)\R]^2\sigma_2(\R)/\sigma_0(\R)$. Lastly, the factor $C$ on
the right, defined as
\beqa
C\equiv\frac{4\pi s^3 N(\sigma'_0,\delta\pko)}{3N(\sigma_0,\delta\pko)}
\int_0^{s}\der r\,3 r^2\,N(\sigma_0,\delta\pko|\sigma'_0,\delta\pko,r)\,
\label{C}
\eeqa
with $s$ equal to the mean separation between the larger scale
non-nested peaks drawn from their mean number density\footnote{This
  must be calculated iteratively, although two iterations, starting
  with $C=1$, are enough to obtain an accurate result.}, is to correct
for the overcounting of background peaks in
$N(\sigma_0,\delta\pko|\sigma'_0,\delta\pko,r)$ as they are not
explicitly required to be non-nested.

The simpler version of the conditional peak number density given in
Section \ref{CUSP} can be readily recovered from the present one by
ignoring the radial dependence of the typical spherically averaged
density profile around peaks, that is taking
$\overline{\delta(r)}=\delta\pko$ and $\Delta \delta(r)=0$.
 
\end{appendix}
\end{document}